# Parallel Implementation of Lossy Data Compression for Temporal Data Sets

Zheng Yuan, William Hendrix[*], Seung Woo Son[+], Christoph Federrath[#], Ankit Agrawal, Wei-keng Liao, and Alok Choudhary

Electrical Engineering and Computer Science Department, Northwestern University
[*]Computer Science Department, University of South Florida
[+]Electrical and Computer Engineering Department, University of Massachusetts Lowell
[#]Research School of Astronomy and Astrophysics, Australian National University
{zys133, ankitag, wkliao, choudhar}@eecs.northwestern.edu, [*]whendrix @usf.edu
[#]christoph.federrath@anu.edu.au, [+]SeungWoo_Son@uml.edu

*Abstract* — **Many scientific data sets contain temporal dimensions. These are the data storing information at the same spatial location but different time stamps. Some of the biggest temporal datasets are produced by parallel computing applications such as simulations of climate change and fluid dynamics. Temporal datasets can be very large and cost a huge amount of time to transfer among storage locations. Using data compression techniques, files can be transferred faster and save storage space. NUMARCK is a lossy data compression algorithm for temporal data sets that can learn emerging distributions of element-wise change ratios along the temporal dimension and encodes them into an index table to be concisely represented. This paper presents a parallel implementation of NUMARCK. Evaluated with six data sets obtained from climate and astrophysics simulations, parallel NUMARCK achieved scalable speedups of up to 8788 when running 12800 MPI processes on a parallel computer. We also compare the compression ratios against two lossy data compression algorithms, ISABELA and ZFP. The results show that NUMARCK achieved higher compression ratio than ISABELA and ZFP.**

I. INTRODUCTION

It is a common practice for parallel applications to checkpoint the intermediate results periodically into files during a long run. The checkpointed files mainly serve two purposes: program restart and post-simulation data analysis. The time stamps associated with these files form the temporal dimension. For large-scale runs, saving checkpoint files to disk can be costly in terms of time and storage space. Modern parallel file systems (PFS) have recently been improved to address I/O performance requirements for large-scale applications. However, shortening the checkpoint time alone is not sufficient when data is shared among groups of researchers and transmitting large size of data online between remote locations is often prohibited because of the sheer volume. A common solution for this problem is to compress the files in situ or post hoc.

Checkpoint files can be considered as an example of temporal data sets. Temporal data sets in this paper refer to a series of files, which record the information of the same spatial space at different time stamps. Traditionally, each file in temporal data sets is individually compressed. Lossless data compression allows data to be reconstructed without losing any precision. However, for high entropy data, lossless compression may fail to achieve a satisfactory compression ratio because common patterns rarely exist in these data sets. Lossy data compression often achieves higher compression ratio, but it faces the challenge to control the error rate.

For most of the scientific simulation data, adjacent iterations are not completely independent from each other. In fact, when one state transitions to the next state, some of the data points share the same changing trends. This property of temporal data sets provides the opportunity to compress data according to the transitions of states. NUMARCK [13] is a lossy data compression algorithm designed based on such concept. By capturing the temporal change ratios between consecutive iterations and encoding changes into indexing space, NUMARCK achieves high compression ratios while guaranteeing the user-defined tolerable error rate. Note the targeted data to be compressed by NUMARCK is the checkpoint data for analysis or visualization, which is usually tolerant of a small degree of errors.

We present a parallel implementation of NUMARCK using MPI. NUMARCK consists of the following computation phases: change ratio calculation, bin construction, data point indexing, lossless index table compression, and file I/O. This paper focuses on discussion of the followings.
1. Parallelization strategy for each phase.
2. A new binning method called top-k binning.
3. File format and layout for storing compressed data.
4. User controllable parameters.
5. Support for partial data decompression.

We evaluated parallel NUMARCK using data sets produced from real scientific simulations: FLASH [14], ASR [15] and CMIP [16]. The experimental results show that the parallel NUMARCK achieves a good scalability and costs small computation time. For instance, it took 2.65 seconds to compress a FLASH variable of size 59GB and achieved a speedup of 1404 when running on 1600 MPI processes. The largest run in our experiment used 12800 MPI processes to compress a FLASH variable of size 472GB and took 3.61 seconds, which corresponds to a speedup of 8788.

We also compare the compression ratios and runtime of NUMARCK against two lossy data compression algorithms, ISABELA and ZFP. In our experiments, NUMARCK outperforms these two algorithms in compression ratios. In addition, Parallel NUMARCK supports partial data decompression. The experimental results show almost a linear relationship between the partial decompression time and the length of the data segment to be decompressed. We also present the results of comparing different binning strategies proposed in earlier work [13]. Among them, top-k binning achieves the best compression performance.

This paper is organized as follows. Section 2 provides an overview of data compression and checkpointing techniques. Section 3 presents NUMARCK, and Section 4 introduces our proposed design and implementation of parallel NUMARCK. In section 5, we evaluate the performance of parallel NUMARCK and Section 6 concludes the paper. We published the code of parallel NUMARCK on Github[31].

## II. BACKGROUND AND RELATED WORK

ZLIB [4] is a commonly used lossless data compression software library. It uses DEFLATE [5], a variation of LZ77 [6], and Huffman coding to compress data. Since ZLIB is a lossless compression algorithm, it may not achieve a good compression ratio for high entropy data.

ZFP [8] is a lossy compression software library to compress floating-point arrays. ZFP splits the data into blocks. Then it applies orthogonal block transformation and encoding on the transformed data. ZFP supports absolute error-bound. However, absolute error-bound sometimes may cause problems when the absolute value of data is small. In addition, it requires the size of each dimension as input parameters.

MCREngine [7] is a data-aware checkpointing system that adopts lossless compression algorithm to compress checkpointing files. In order to achieve high compression ratio, it aggregates data with similar meaning from multiple application processors. Then, it uses lossless compression to compress the aggregated data. According to their result, it achieves a compression ratio of 1.5-4 on ALE3D [17], Cactus [18][19] and Enzo data sets [20]. The drawback of MCREngine is that it needs additional metadata to represent the meaning of data. Moreover, if the data itself is random, MCREngine cannot achieve a high compression ratio.

Sasaki, Naoto, et al [9] propose a lossy data compression method using Haar wavelet transformation in checkpointing system. According to their results, the overall checkpoint time reduced by 81% for a production climate application, NICAM [21]. The compression ratio is about 6.5, while the average error rate was 1.2%. However, their solution cannot control the error, which may cause some data points to change too much after reconstruction.

ISABELA [10] applies a pre-conditioner to high entropy data along spatial resolution to achieve an accurate fitting model within user defined error-bound. To get better fitting performance, it sorts the data in a non-decreasing order. According to the experimental results in [10], ISABELA achieves higher compression ratios compared to other lossy compression techniques such as Wavelet compression.

Son, Chen, et al [22] describe several lossy compression algorithms that radically change how checkpoint data is stored with tunable error bounding mechanisms. It predicts lossy compression to be a promising way to reduce checkpoint overheads without compromising the quality of datasets that scientific simulation operates on.

Gilchirst and Cuhadar [24] present two different parallel implementations (data parallel and task parallel) of BWT-based lossless data compression algorithms. To implement data parallel compression, the entire data file is split into small chunks. Each chunk is compressed individually and written to disk in order. To implement task parallel compression, different threads or processes respond to different stages of the compression algorithm. Experimental results show significant speedup for the parallel data approach.

Patel, Zhang, et al [25] present a parallel implementation of bzip2-like lossless data compression scheme for GPU. They parallelize three main stages of compression pipeline: Burrows-Wheeler transform (BWT), move-to-front transform (MTF), and Huffman coding. The experimental results show that their implementation is slower than that of bzip2.

Di and Cappello [29] optimize the error-bounded HPC data compression. The idea of their work is to fit/predict the successive data points with the best-fit selection of curve fitting models. The experimental results show that the compression ratio of the proposed compressor ranges from 3.3 to 436. Compared to other compression methods, their compressor achieves a comparable compression time and 50% to 90% shorter decompression times.

## III. NUMARCK OVERVIEW

NUMARCK consists of three computation phases: change ratio calculation, binning, and indexing. We use the following symbols to help describe NUMARCK.

$B$ : number of bits to index a data point

$E$ : user defined tolerance error bound

$n$ : total number of data points

$D_{i,j}$ : the $j^{th}$ data point in input data set $D$ at the $i^{th}$ iteration

$R_{i,j}$ : the data point reconstructed from NUMARCK that corresponds to $D_{i,j}$

$\Delta D_{i,j}$: change ratio of $D_{i,j}$ from the previous iteration ($i$-$1$).

$$\Delta D_{i,j} = \frac{D_{i,j} - D_{i-1,j}}{D_{i-1,j}} \quad (1)$$

$CR$: the compression ratio.

$$CR = \frac{original\ file\ size}{compressed\ file\ size} \quad (2)$$

$ME$: the mean error rate.

$$ME = (\sum_{j=1}^{n} \left| \frac{D_{i,j} - R_{i,j}}{D_{i,j}} \right|)/n \quad (3)$$

*A. Phase 1: Change ratio calculation*

Based on equation *(1)*, NUMARCK calculates the element-wise change ratios for data sets at time stamp *i* from its previous time stamp *i-1*. The time complexity of this phase is *O(n)*. Change ratios are stored in a float-point array of size *n*.

*B. Phase 2: Bin construction*

NUMARCK splits the range of change ratios into bins. The change ratios falling in the same bin are approximated by the bin center. This strategy allows us to use the bin index, an integral value, to represent all the data points whose change ratios fall into the same bin. Three binning strategies were presented in [13]: equal-width binning, log-scale binning, and k-means data clustering based binning. Different binning strategies produce different bin centers, resulting in different compression ratios. The equal-width binning method evenly splits the range of change ratio into $(2^B-1)$ bins. The log-scale binning method calculates the bin width based on a log scale. The k-means method applies the k-means data clustering algorithm to group change ratios into $k = (2^B-1)$ bins. The results indicated that the k-means strategy outperforms the other two. The time complexity of this phase for equal-width and log-based strategies are both $O(n)$. For the k-means binning, the complexity is $O(n * 2^B * I)$, where *I* is the number of k-means iterations. Output of this phase contains two arrays: bin centers, a float-point array of size $2^B$, and bin indices, an integer array of size *n*.

*C. Phase 3: Indexing*

In this phase, data points falling into the $(2^B-1)$ bins selected from the previous phase will be indexed by the corresponding bin IDs. We refer these data points as compressible data. Data points whose change ratios are not in any of the selected $(2^B-1)$ bins are marked as incompressible data. All incompressible data points are indexed by the last index value, $2^B-1$ and their values are stored as is in a separate array called incompressible data array. Compressible data that fall into the same bin are represented by the bin's ID. At the end of this phase, all data points are converted to an index of size *B* bits and stored in the index table of size *(n*B)* bits. The time complexity of this phase is *O(n)*.

*D. File I/O*

NUMARCK writes the compressed data into files using netCDF format [28]. The bin centers, index table and incompressible data will be saved as array variables in the netCDF file. Metadata describing these variables, such as number of data elements, number of centers etc., is defined as netCDF attributes.

*E. Decompression*

To reconstruct a data point, NUMARCK first finds its index from the index table. If it is marked as an incompressible data point, its original value can be retrieved from the incompressible data array. If it is compressible, we reconstruct it by multiplying its value in previous iterations by the change ratio of its bin center based on equation (4).

$$R_{i,j} = \Delta D_{i,j} * R_{i-1,j} \qquad (4)$$

IV. PARALLEL NUMARCK

We describe our parallelization strategy for NUMARCK in distributed computing environments. Without loss of generality, we assume data points are evenly distributed among all processes.

*A. Parallel change ratio calculation*

Based on equation (1), all processes calculate the point-wise change ratios for the local data set. This phase can be done in parallel without inter-process communication. The resulting change ratios are saved in an internal buffer array of float-point type. During the calculation, we also track the minimum and maximum change ratios. A collective call to *MPI_Allreduce* gives the global minimum and maximum among all processes. The range of change ratios is subsequently used in the bin construction phase.

*B. Parallel bin construction*

In this section, we first propose a new binning strategy named *top-k binning* and present its parallelization. Then, we describe a method for selecting the optimal *B* value. The parallelization of the equal-width binning, log-scale binning and k-means binning are described at the end.

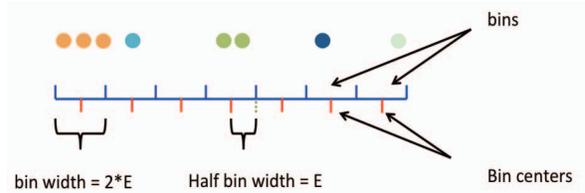

Figure 1. An example of top-k binning strategy

*1) Top-k binning strategy*

*Top-k binning* first divides the change ratio range into bins of width *2E* each. In the next step, a histogram is constructed by counting the number of change ratios falling into bins. The last step is to select top k bins that contain the most number of change ratios. Figure 1 depicts the data points and bins. Colored dots represent the change ratios. Top-k binning first splits the range of change ratio into bins, which are represented by blue cells. The width of each bin is *2E*. Red ticks denote the bin centers. It then counts the number of change ratios falling in each bin. Data points whose change ratios are located in the same bin share the same color. Note in our case, $k = (2^B-1)$ is the size of index table that stores the compressible data points. To implement top-k binning using MPI, a call to *MPI_Allreduce* combines the local histograms into a global histogram on all processes. A *top-k selection* algorithm is used to find the k bins from the global histogram that contain the largest number of change ratios.

*2) Selecting the top k bins*

Since the value of $B$, or $k$ as $k = (2^B-1)$, determines the sizes of the bin center table, index table and incompressible data table, selecting its value must be done carefully to achieve a good compression ratio. In [13], $B$ is user defined. Note that in the top-k strategy, the width of each bin is $2E$. Given the generated histogram described above, we can calculate the percentages of incompressible data for different $B$ values. Consequently, NUMARCK can estimate the compressed data sizes and select a $B$ value that results in the highest compression ratio.

Given a $B$, there are $2^B$ different values in the indexing space. Thus, the space available to store the bin center array is $(2^B*L)$ where $L$ is the byte size of original data point's data type, e.g. 4 for float and 8 for double. The space required to store the index table is $(n*B/8)$ bytes, because each index is of size $B$ bits and the index table contains $n$ elements. The size of incompressible data array is represented as $(n*\alpha*L)$ where $\alpha$ is the incompressible data ratio, defined as

$$\alpha = \frac{n - \text{number of compressible data points}}{n} \quad (5)$$

Therefore, the estimated compressed file size becomes

$$file\_size(B) = 2^B * L + (n * \frac{B}{8}) + n * \alpha * L \quad (6)$$

Since each process has the same copy of the global histogram, no inter-process communication is needed in this phase. All processes calculate *file_size(B)* for a range of $B$ values and select the one that has the lowest *file_size(B)*.

*3) Parallel implementaion of other binning strategies*

To implement equal-width binning in parallel, each MPI process selects local minimum and maximum change ratios. Using MPI_Allreduce(), each process then gets the global minimum and maximum change ratios and evenly divides the global change ratio range into $2^B-1$ chunks. The centers of these chunks are selected as the bin centers. The implementation of log-scale binning is similar to equal-width binning except each process calculates the maximum and minimum log value of the change ratios. NUMARCK uses the MPI k-means package provided by Liao [30] to parallelize the k-means binning.

*C. Parallel indexing*

NUMARCK splits the index table into blocks of same size and compresses each block using the ZLIB to further compress the index table. When $n >> 2^B$, the index table may contain many repeated index values, since indexing space is limited to $2^B$. This motivates us to compress the index table using a lossless compression algorithm.

Our parallel indexing strategy is deigned to support "partial data decompression". NUMARCK splits an index table into blocks of equal size. Each block is compressed separately by ZLIB. Therefore, when decompressing a part of the data, instead of decompressing the whole index table, we only need to the index table blocks that contain the desired data points. Since the length of compressed block is unpredictable, we use an offset table to store the starting offsets of each block. Note that an index table is a bit array, but an index table block is aligned in bytes to be compressed by ZLIB. Therefore, there may exist several unused bits at the end of each index block. In addition, to calculate the offset of an incompressible data point in an incompressible data table, the number of incompressible data points before each block needs to be saved in another offset table.

In the parallel indexing phase, each process calculates the indices for local data points. This can be done without inter-process communication. Assuming the block size of 256KB, each block can store up to $\lfloor 256KB/8/B \rfloor$ indices. Each process first calculates the boundaries of blocks locally and then calls *MPI_Scan* to obtain the global block boundaries and calculates the "owners" of the blocks. The block assignment is done in a contiguous chunk fashion. Based on the global boundaries, all processes transfer indices with their left or right neighbors using *MPI_Send and MPI_Recv*. Once the data transfer is complete, all processes bit-write indices to local bits buffers (index table blocks). As for the bit operation, NUMARCK first stores the indices as 4- or 8-byte integers. Then, it bit-copies the B least significant bits of the integer to the corresponding index table entry. At the end of this phase, all processes call ZLIB to compress their assigned index blocks.

*D. Parallel file write*

Using PnetCDF library [22], NUMARCK stores the compressed data in the netCDF file format. Bin centers and incompressible data are saved in 1D arrays of float type. All index table blocks stack up into a 'big index table'. The 'big index table' is stored as an array of byte type. Two more integer arrays are added for the offset tables. Offsets of incompressible data for each block are stored in another array of long long type. NUMARCK allows multiple compressed variables stored in one netCDF file. Figure 2 shows the netCDF file header information of a NUMARCK file. In this example, there is only one variable, named 'UU' from the ASR data set. Detailed information of this variable can be found in the evaluation section.

The attributes of UU_info includes the number of data points, the number of bin centers, and the number of indices per block. Array UU_bin_centers contains the bin centers of 'UU'. In this example, there are 8192 bin centers ($B=13$) used to compress 'UU'. Following these two variables are the offset arrays of the index table and the incompressible data array. In this example, there are six index blocks. The indices of compressible data and the incompressible data are stored in UU_index_table and UU_incompressible_table, respectively.

V. PERFORMANCE EVALUATION

We evaluate NUMARCK on two machines: a Linux machine with 2 quad-core CPUs, Xeon E5-2407 2.20 GHz (32 GB memory) and SuperMUC, a parallel computer at the Leibniz Supercomputing Centre. For small data sets(Sedov,

Table 1. A summary of test data sets

| Name of data set | Domain application | Size per iteration | Data type | Variable dimension | Variable size | Number of variables per file | Number of iterations |
|---|---|---|---|---|---|---|---|
| Sedov | FLASH | 15MB | Double | 165*32*32*1 | 1.3MB | 10 | 40 |
| Stir-1 | FLASH | 3.7GB | Float | 64*157*157*157 | 945MB | 4 | 100 |
| Stir-2 | FLASH | 296GB | Float | 1024*314*314*157 | 59GB | 5 | 100 |
| Stir-3 | FLASH | 2.4TB | Float | 8192*314*314*157 | 472GB | 5 | 100 |
| ASR | ASR | 103MB | Float | 29*320*320 | 11MB | 10 | 80 |
| CMIP | CMIP3 | 19GB | Float | 42*2400*3600 | 1.4GB | 13 | 6 |

```
dimensions:
    UU_info = 1 ;
    UU_bin_centers = 8192 ;
    UU_block_table_dim = 6 ;
    UU_index_table_dim = 5663344 ;
    UU_incompressible_table_dim = 92236 ;
variables:
    int UU_info(UU_info_dim) ;
        UU_info:total_data_num = 3758400 ;
        UU_info:bin_centers_number = 8192 ;
        UU_info:elements_per_block = 645277 ;
    float UU_bin_centers (UU_bin_centers_dim) ;
    int UU_index_table_offset(UU_block_table_dim) ;
    int UU_incompressible_table_offset(UU_block_table_dim) ;
    byte UU_index_table(UU_index_table_dim) ;
    float UU_incompressible_table (UU_incompressible_table_dim) ;
```

Figure 2. Header of a NUMARCK compressed file in netCDF format..

Stir-1, ASR and CMIP), we ran NUMARCK sequentially on the Linux machine. For large datasets(Stir-2, Stir-3), we ran parallel jobs on SuperMUC. All our parallel experiments ran on the 'Thin Nodes Phase 1' cluster in SuperMUC. 'Thin Node Phase 1' consists 9216 nodes (147,456 cores) and each node has 32GB memory.

The datasets used in our experiments are data collected from FLASH [14], ASR [15] and CMIP [16]. Table 1 summaries all test data sets that are used in this paper. FLASH [14] is a block-structured, multi-physics, hydrodynamic code that solves the compressible Euler equations on a block-structured adaptive mesh. FLASH divides the problem domain into blocks with a fixed number of cells. A block is a three-dimensional array with $(N+4)^D$ cell, consisting of N active cells and 2+2 guard cells in each spatial dimension D on both sides to hold information from its neighbors (used for communication).

We use the output of two FLASH setups, the Sedov-Taylor bastwave test (sedov) and the turbulence stirring test (stir). As for sedov data, we set the block size to be 32×32 cells and run the FLASH_sedov 2D simulation test for 40 iterations. The size of each checkpoint file is 15 MB per iteration. Stir [26], [27] is a simulation setup of driven turbulence of interstellar gas. We take 11 snapshots of simulations that vary in the total number of grid cells. We have simulations with $628^3$, $1256^3$, $2512^3$, $5024^3$, and $10048^3$ grid cells. Each of these files is 3.7GB, 37GB, 296GB, 2.4TB, and 19TB, respectively, each containing 5 variables of single-precision float type for visualization and analysis purposes. A turbulent system has intrinsically high entropy and is thus very hard to compress, so we use it here as a challenging test case for NUMARCK. We note that the initial conditions or states after only a few time steps of Stir will be relatively easily compressible because the gas only starts to become fully turbulent in the first few iterations. As shown in [26], after two turbulent crossing times (2T), the turbulence evolves into a fully developed turbulent field, which is hard to compress. Here we use 11 checkpoint files from 2T to 3T, spaced by 0.1T. In this paper, we use $628^3$, $2512^3$, $5024^3$ as test data sets. We use the name of Stir-1, Stir-2 and Stir-3 to refer these three data respectively.

The Arctic System Reanalysis (ASR) [15] provides atmosphere data of the Arctic. The data consists of 29 pressure levels, 27 surface and 10 upper air analysis variables, 74 surfaces, 16 upper air forecast variables, and 3 soil variables. The format of the ASR files is netcdf-4. The files we use contain atmosphere data for 10 consecutive days, 80 iterations, from Jan $1^{th}$ to Jan $10^{th}$ in 2004. The size of ASR files is 103MB per iteration.

CMIP [16], the Coupled Model Intercomparison Project, is the analog of AMIP for global coupled ocean-atmosphere general circulation models. We use CMIP3 output. CMIP3 output from coupled ocean-atmosphere model simulations of 20th - 22nd century climate is being collected by the PCMDI in support of research relied on by the $4^{th}$ Assessment Report (AR4) of the Intergovernmental Panel on Climate Change (IPCC). The size of CMIP3 is 19GB per iteration. We downloaded six iterations of CMIP data for our compression tests.

*A. Parallel NUMARCK evaluation*

To evaluate the runtime and scalability of the parallel NUMARCK, we run NUMARCK to compress *velx* variable in Stir-2 in parallel using up to 1600 cores and the same variable in Stir-3 data using up to 12800 cores. *Velx* represents the turbulent velocity components along x-axis. The sizes of *velx* are 59GB and 472GB respectively. Other data sets are not used in this sub-section because of their small size. Top-k binning strategy is used in this sub-section. I/O time is excluded.

Table 2 illustrates the parallel runtimes of NUMARCK on Stir-2 and Stir-3 data sets. The parallel run time decreases dramatically as the number of processes increases in both cases. It takes only 2.655 seconds to compress a variable of size 59GB and 3.610 seconds to compress a variable of size 472GB using 1600 and 12800 CPU cores, respectively. The results show NUMARCK is a fast parallel data compression algorithm for large data sets. Figures 3 and 4 show the speedup curves for both of the data sets. The speedups for Stir-2 data set are almost linear. For Stir-2 data set, NUMARCK achieves a speedup of 1404 with 1600 cores. For Stir-3 data set we achieved speedup of 8788 using 12800 cores. Both speedup charts indicate NUMARCK scales well along with the number of processes.

Figures 5 and 6 plot the timing breakdown in percentages for individual phases. We split the indexing phases into three sub-phases: assigning index (assigning an integer index to each data point), index alignment (exchanging indices with neighbor processes according to the partitioning boarder of index table), and bits packing (bit-writes the integer index into bits buffer.) We observe that in all cases the inter-process communication cost for index alignment is very little. In our experiments, we set the block size to 1MB. Each MPI process exchange only 2MB (at most) data with their neighbors, resulting a small communication cost.

Table 2. Parallel run time (in seconds) for Stir-2 and Stir-3 data sets

| Num of cores | Runtime of Stir-2 | Num of cores | Runtime of Stir-3 |
|---|---|---|---|
| 320 | 11.650 | 3200 | 9.914 |
| 480 | 7.827 | 4800 | 7.289 |
| 640 | 5.874 | 6400 | 5.637 |
| 800 | 4.798 | 8000 | 4.813 |
| 960 | 4.057 | 9600 | 4.529 |
| 1120 | 3.534 | 11200 | 3.996 |
| 1280 | 3.222 | 12800 | 3.610 |
| 1440 | 2.943 | | |
| 1600 | 2.655 | | |

The rest five phases (ZLIB compression, indexing, change ratio calculation, assign index and bits packing) take more than 98% of the runtime. The scalability of these phases dominates the overall speedup. Note that there is no network communication cost in the phases of ZLIB compression, change ratio calculation assign index, and bits packing. In addition, the data need to be proceed in these four phases are evenly divided among all MPI processes. So, these phases are expected to be very scalable. As described in section IV, indexing phase (top-k binning) contains two parts that impacts the scalability of the binning phase: top-k selection and *MPI_Allreduce()*. Each MPI process runs the top-k selection on the same copy of the global histogram. The top-k selection is regarded as a serial part. Thus, using more processes cannot reduce the runtime top-k selection. In addition, MPI global reduction becomes costly as the number of cores increases, which degrades the overall scalability. To further investigate the impact of the cost of top-k selection and MPI_Allreduce() on binning phase, table 3 shows the runtime percentages for these two parts (over the whole binning phase) for the smallest and largest scale runs on both Stir data sets. Although top-k selection is serial, its run time occupies less than 1.5% of the binning phase, which has negligible impact on the overall speedup. However, the *MPI_Allreduce()* occupies a very large proportion of the runtime, especially for large scale runs, which thus worsen the speedup of the binning phase.

The speedup breakdowns of each phase are plotted in figure 7 and 8. We observe that all phases scale well except the binning phase. For Stir-2, the speedup of binning phase levels off after 960 cores. As for the Stir-3, the binning phase does not scale from the beginning. The entire binning phase takes only 0.8-1.5 seconds. Because of the short binning time, *MPI_Allreduce()* becomes relatively significant and thus dominates in large scale runs as we discussed previously. This causes the limited speedups in the binning phase. Since the binning phase takes only a small fraction of the total runtime and all other phases scale much better, NUMARCK still achieves good overall speedups.

Table 3. Runtime percentages of top-k selection and MPI_Allreduce()

| Stir-2 | | | Stir-3 | | |
|---|---|---|---|---|---|
| | 320 cores | 1600 cores | | 3200 cores | 12800 cores |
| Top-k selection | 1.4% | 1.2% | Top-k selection | 1.0% | 0.8% |
| MPI_Allreduce | 5.0% | 67.6% | MPI_Allreduce | 40.5% | 81.6% |

### B. Compression ratios of NUMARCK

We compare the compression ratios achieved by NUMARCK to two well-known lossy data compression algorithms, ISABELA and ZFP. In order to evaluate NUMARCK and compare it with other lossy compression algorithms, we use the error-threshold 0.1%. Note that the number of MPI processes does not influence the compression ratio of NUMARCK. The user-defined error thresholds used in ISABELA and ZFP are also set to 0.1%. Since ZFP only provides the absolute error bound control, to achieve fair compression, we set the absolute error bound of ZFP to the mean value of the test data multiplied by 0.1%. ISABELA and ZFP are serial compression algorithms. They cannot handle large data sets, i.e. Stir-2 and Stir-3, because the memory size on one node is only 32GB. In this section, we only evaluate the compression performance on Sedov, Stir-1, ASR and CMIP data. We select one variable per data set as test data in this sub-section. We select the *ener* variable in the Sedov simulation for compression, which is a 4-dimensional double-float array. The *ener* variable represents a discretized approximation of specific energy in a grid cell, computed as the quotient of total energy and total mass of the cell, where total energy includes thermal energy and macroscopic kinetic energy. The dimension of *ener* is: 165*32*32*1. We select the *dens* variable in the Stir-1 simulation. This variable represents the density of the turbulent gas. It is a single-float array. The dimension is 64*157*157*157. We select the *UU* variable in the ASR data set. *UU* is a single-float array, which represents the wind speed along the x-axis in the grid. The dimension of *UU* is 29*360*360. We select the UVEL variable in the CMIP as test data. UVEL represents the current velocity along the curvilinear grid. This variable is a 42*3600*2400 dimension single-float array.

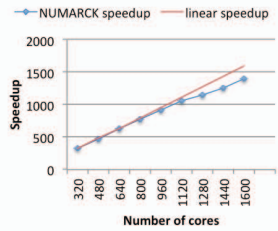
Figure 3. Stir-2 speedup of NUMARCK
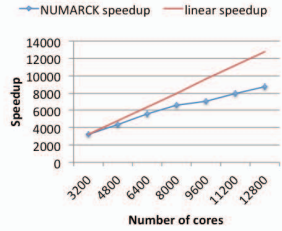
Figure 4. Stir-3 speedup of NUMARCK
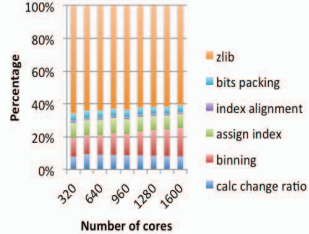
Figure 5. Stir-2 runtime breakdown
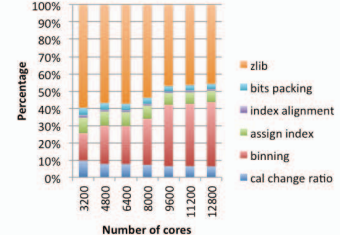
Figure 6. Stir-3 runtime breakdown

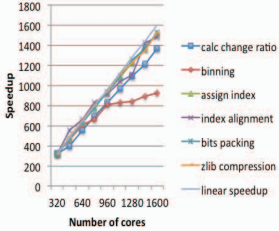
Figure 7. Stir-2 speedup breakdown
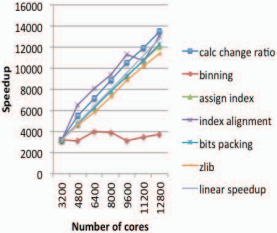
Figure 8. Stir-3 speedup breakdown
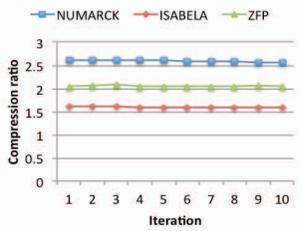
Figure 9. Stir-1 compression ratio
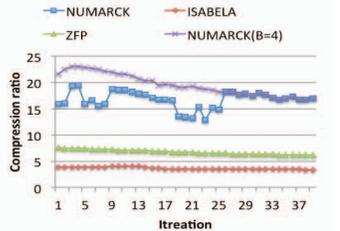
Figure 10. Sedov compression ratio

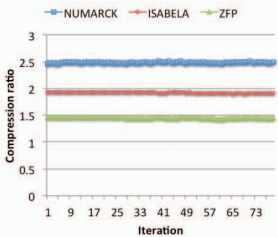
Figure 11. ASR compression ratio
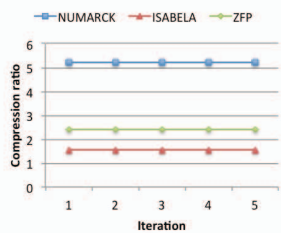
Figure 12. CMIP compression ratio
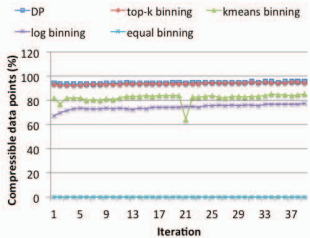
Figure 13. Sedov compressible data points
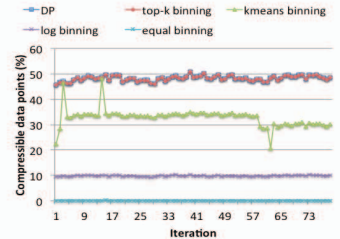
Figure 14. ASR compressible data points

Figure 9-12 show the compression ratios achieved by NUMARCK, ISABELA, and ZFP on Stir-1, Sedov, ASR and CMIP data sets. In this experiment, only the top-k binning strategy is used for NUMARCK. We observe that the compression ratios of NUMARCK are always better than ISABELA and ZFP. The high compression ratios of NUMARCK can be explained by the low incompressible data ratios. Table 4 shows the average incompressible data ratios of all data sets. In other words, most of the data points are compressible and can be represented by indices of size a few bits, which results in high compression ratios in NUMARCK. The reason for the low incompressible ratios can be explained as follows. For temporal data sets, a lot of data points share the same changing trends (similar change ratios), and thus fall into a small number of bins.

Table 4. Average incompressible ratios of all data sets

| Stir(3.7GB) | Sedov | ASR | CMIP |
|---|---|---|---|
| 1.96% | 4.45% | 4.45% | 2.04% |

We now use an example to analyze how higher compression ratios were achieved by NUMARCK for CMIP data. Note that only 2% of the data points are incompressible in the data set. The index length of CMIP is 12, which means almost all single-float data points (32 bits) are converted to 12 bits. In addition, the compression ratio of the CMIP index table is 2.2. So, the compression ratio of CMIP can be roughly calculated as $\frac{32}{(12/2.2)+2\%*32} = 5.2$. Although the incompressible ratios of Stir-1 and ASR are similar to that of CMIP, the compression ratio of Stir-1 and ASR is not as high as that of CMIP. The reason is the index table compression ratios of Stir-1 and ASR are about 1.1. As for the test data, the index table occupies more than 95% of space in the compressed files. Therefore, the ZLIB compression ratio greatly influences the compression ratio of NUMARCK. The more repeated indices exist in the index table, the higher compression ratios ZLIB achieves.

We observe an interesting curve of the NUMARCK compression ratios for Sedov: the compression ratios fluctuate significantly. This phenomenon is caused by the imprecise index length selection and the index lengths that are not consistent among MPI processes, leading to fluctuated file size. We also plot the compression ratios for Sedov data using B=4 (purple line) which show no fluctuation. We will discus this phenomenon in sub-section D.

The compress time and the decompress time of NUMARCK, ISABELA, and ZFP are shown in table 5 and 6. We observe that both compression and decompression times of ISABELA are longer than both ZFP and NUMARCK in all cases. ZFP always achieves the shortest compression time. NUMARCK runs two times slower than ZFP for Stir-1, Sedov and ASR data sets and 1.2 times slower for CMIP data set. As for the decompression time, NUMARCK is the fastest one. It is because each data point is reconstructed by a multiplication according to equation (4), which only requires $O(n)$ time. Although NUMARCK is not as fast as ZFP in the compression, it produces a higher compression ratio.

## C. Partial decompression

NUMARCK supports partial decompression. To evaluate the partial decompression performance for all data sets, we randomly select a starting point and chunk 20%, 40%, 60% and 80% of the data separately. The decompression times of these segments are listed in table 7. The decompression times of the whole data sets (100%) are also included. Table 7 shows an almost linear relationship between the segment lengths and the decompression times on Stir-1, ASR and CMIP data. However, the partial decompression times of Sedov are not much different. This is because there is only one block in Sedov data and the index table block decompression is the most time consuming part in the decompression phase. So, for any proportion of the partial decompression, the single index block has to be decompressed. This results in similar partial decompression times of the Sedov data.

## D. Evaluation of top-k binning strategies

In this sub-section, we show that top-k binning strategy outperforms other binning strategies and top-k binning is able to auto-select a good index length.

Different binning strategies select different bin centers, which results in different incompressible ratios. In this sub-section, we show a dynamic programming (DP) solution for top-k binning in terms of the compressible ratio. We prove that no other binning strategies can cover more data points than the DP solution. We compare the number of compressible data points and the runtime of all the binning strategies. Experimental results show that, while the runtime is short, the top-k binning strategy covers almost the same amount of data points as the DP solution does.

Theoretically, the binning problem can be described as the following:

**Input**: $n$ data points and their values, bin width $W$, an integer $k=2^B$.

**Output**: The maximum number of data points that can be covered using $k$ bins with length of $W$ each.

Note that the binning problem can be solved in polynomial time using dynamic programming. The sub-problem of the DP solution can be defined as 'OPT(i,j) is the largest number of data points from points i to N in sorted order that can be covered with j bins.' We have, OPT(i,j) = max(OPT(i+1,j), OPT(i+c(i),j-1)+c(i)). c(i) is defined as the number of data points covered by the interval [Value$_i$, Value$_i$+W]. The initial states are OPT(N+1, $\forall$ j) = 0, OPT($\forall$ i, 0) = 0. Figure 15 shows the algorithm of the DP solution.

We now prove the correctness of this DP solution. Consider for the i$^{th}$ data point, we are computing OPT(i,j). We need to decide if we should put the i$^{th}$ data in a bin or not. If we do not put the data in a bin, then the best solution is to use j bins to cover the data belonging to [i+1,N], which is OPT(i+1,j). Otherwise, we cover the data using a bin and calculate the number of elements covered by this bin. Then, we add the number of covered data point using j-1 bin for the rest of the data, i.e. OPT(i+c(i),j-1)+c(i).

However, this dynamic programming solution requires $O(n*2^B)$ memory space, which is impractical for parallel data compressions. For instance, to compress 1GB of data using B=10, requires 1TB of memory.

```
OPT[∀ i][0] = 0;
OPT[N+1][ ∀ j] = 0;
for(i=N to 1)
    for(j=1 to k)
        OPT[i][j] = max(OPT[i+c(i)][j]+c(i),OPT[i+1][j-1]);
```

Figure 15. The algorithm of dynamic programming for binning strategies

Table 5. Compression time of NUMARCK, ISABELA and ZFP (seconds)

| Data set | NUMARCK | ISABELA | ZFP |
|---|---|---|---|
| Stir (3.7GB) | 64.87 | 1007.31 | 27.74 |
| Sedov | 0.037 | 0.378 | 0.022 |
| ASR | 1.31 | 17.51 | 0.47 |
| CMIP | 65.1 | 1170.9 | 54.3 |

Table 6. Decompression time of Numarck, ISABELA and (seconds)

| Data set | NUMARCK | ISABELA | ZFP |
|---|---|---|---|
| Stir (3.7GB) | 14.267 | 104.210 | 44.947 |
| Sedov | 0.012 | 0.095 | 0.030 |
| ASR | 0.334 | 1.615 | 0.731 |
| CMIP | 15.77 | 151.03 | 73.800 |

Table 7. Partial decompression time (in seconds)

| Data size | Stir-1 | Sedov | ASR | CMIP |
|---|---|---|---|---|
| 20% | 2.835 | 0.009 | 0.068 | 3.261 |
| 40% | 5.471 | 0.010 | 0.135 | 6.506 |
| 60% | 8.207 | 0.011 | 0.200 | 9.723 |
| 80% | 10.966 | 0.011 | 0.268 | 12.90 |
| 100% | 14.267 | 0.012 | 0.334 | 15.77 |

Figures 13 and 14 compare the number of compressible data points covered by different binning strategies and the dynamic programming solution using Sedov and ASR data. We use the same data sets as in the previous sub-section. The results of other data sets are not shown since the DP solution requires too much memory and the k-means binning takes too much time to run. We set B=8 for Sedov and B=14 for the ASR data. (Data points with change ratios below |E| are excluded.)

Table 8. Binning strategies runtime (milliseconds)

| Data sets | DP | top-k | kmeans | log-bin | equal bin |
|---|---|---|---|---|---|
| Sedov | 92.81 | 18.15 | 645.26 | 4.95 | 0.77 |
| ASR | 13.94 | 0.44 | 45.89 | 0.36 | 0.12 |

Because of the large change ratio range, the bins of equal binning are wide. Only a few data points are located near the center of a bin. This leads to lower compressible ratios of equal binning. Log binning performs better than equal binning but worse than k-means binning. We observe that top-k covers almost the same amount of change ratios as the DP solution does in both data sets.

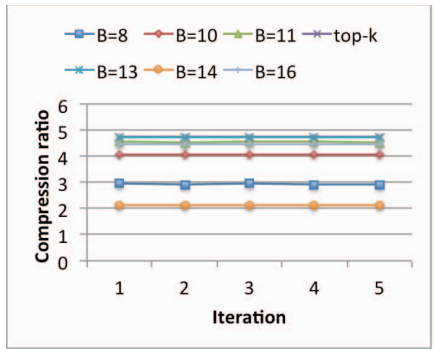

Figure 16. compression ratios of ASR data for different B.

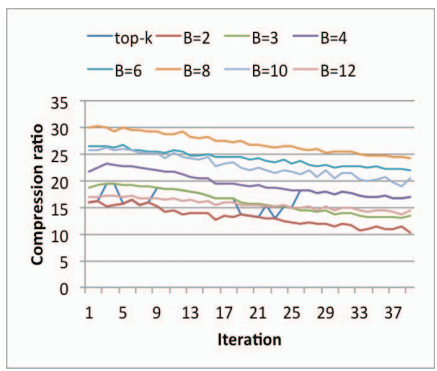

Figure 17. compression ratios of Sedov data for different B.

Table 8 shows the runtime of different binning strategies and the DP solution. As we can see, k-means binning runs slowest, while equal binning runs the fastest. Top-k binning is much faster than the DP solution.

Table 9. average zlib compression ratio

| ASR | | Sedov | |
|---|---|---|---|
| B | Average ZLIB compression ratio | B | Average ZLIB compression ratio |
| 8 | 2.56 | 2 | 9.65 |
| 10 | 1.48 | 3 | 8.92 |
| 12 | 1.36 | 4 | 9.79 |
| 13 | 1.26 | 6 | 9.87 |
| 14 | 1.27 | 8 | 11.00 |
| 15 | 1.24 | 10 | 10.08 |
| 16 | 1.32 | 12 | 11.34 |

An interesting question is: can the top-k binning method auto-select an optimal B value to maximize the compression? We first run NUMARCK using top-k binning strategy with a set of different B values to obtain the compression ratios. (Note that in order to do this, we set the B values by hand and disable the auto-selection of B.) Then, we check whether the B value auto-selected by the top-k binning achieves the highest compression ratio. We use CMIP, ASR, Sedov and Stir-2 as test data sets. In order to save computational recourses, we compress the Stir-2 data every five iterations. (The result of other data sets is not shown due to the space limitation. They behave almost the same as ASR data.)

Figure 16 shows the compression ratios using different B for ASR data. Top-k auto-select B equals to 14 in this case. As is shown in figure 16, if B is too big (B=16) or too small (B=8), NUMARCK does not get the highest compression ratio. NUMARCK achieves the highest compression ratio when B=12. However, top-k binning strategy selects 14 as the index length. This is caused by ZLIB compression. The average ZLIB compression ratio for ASR data is shown in the first two columns of table 9. We observed that the ZLIB compression ratio is around 1.3 for ASR data. However, when choosing the index length, top-k binning compares the file lengths of different B without considering the ZLIB compression. Thus, top-k may mis-predict the file length and set the index length to a non-optimal value. Although top-k does not select the optimal index length, the compression ratios of B=14 are very close to the optimal B as is shown in figure 16.

Figure 17 shows the compression ratios using different B for Sedov data. In this case, top-k sets B to two, three and four for different iterations. However, the optimal B is 8 and the compression ratio is almost twice bigger than that of top-k auto-selection. The low compression ratios are mainly caused by the high ZLIB compression ratios, as is shown in the last two columns in table 9. The ZLIB compression ratio of index table is around 10. This results in significant differences between the compressed file length and the estimated file length. Consequently, top-k binning selects an inappropriate index length, which results in low compression ratios. To understand the reason for the high ZLIB compression ratios for Sedov data, we investigate the distribution of change ratios. We found that there are 80% of the Sedov data points, where change ratios are smaller than the error bound. The change ratios of these tiny changed or unchanged data points fall in the same bin and are represented by the same index, which causes high ZLIB compression ratios of the index table.

## VI. CONCLUSION

NUMARCK is a lossy data compression algorithm designed for temporal datasets such as checkpoint files produced for analysis purpose by large-scale simulation programs. It captures the distribution of pairwise change ratios of data points between two adjacent time steps and performs an in-situ data compression through approximation, while guaranteeing the user-defined element-wise error bound.

In this paper, we presented a design and parallel implementation of NUMARCK using MPI. The experimental results show scalable speedups up to 1404 using 1600 cores for Stir-2 and 8788 using 12800 cores for Stir-3. We compared NUMARCK against two lossy data compression algorithms, ISABELA and ZFP. Our experimental results show that NUMARCK performs better than ISABELA and ZFP on the temporal data sets in terms of the compression ratios and run time. We also evaluated the partial decompression time of NUMARCK. The result shows almost a linear relationship between the partial decompression time and the length of the data segment. In addition the three binning strategies presented in our earlier work, we propose a new binning strategy, called top-k. Our evaluation shows that, while the runtime of top-k binning is very short, it yields almost the same result as the DP solution does. We showed that the top-k binning strategy outperforms previously proposed strategies and it enables an auto-selection for a good index length that results in a high compression ratio.

ACKNOWLEDGEMENT

This work is supported in part by the following grants: NSF awards CCF-1029166, IIS-1343639, CCF-1409601; DOE awards DE-SC0007456, DE-SC0014330; AFOSR award FA9550-12-1-0458; NIST award 70NANB14H012; DARPA award N66001-15-C-4036. C.F. acknowledges funding provided by the Australian Research Council's Discovery Projects (grants DP130102078 and DP150104329). C.F. further acknowledges supercomputing time provided by the Jülich Supercomputing Centre (grant hhd20), the Leibniz Rechenzentrum and the Gauss Centre for Supercomputing (grants pr32lo, pr48pi and GCS Large-scale project 10391), the Partnership for Advanced Computing in Europe (PRACE grant pr89mu), the Australian National Computational Infrastructure (grant ek9), and the Pawsey Supercomputing Centre with funding from the Australian Government and the Government of Western Australia. The simulation software FLASH was in part developed by the DOE-supported Flash Center for Computational Science at the University of Chicago.